\documentclass[twocolumn,showpacs,preprintnumbers,amsmath,prl]{revtex4}
\usepackage{epsfig}
\usepackage{amsmath}
\usepackage{graphicx,natbib}
\begin{document}

\title{Growing Directed Networks: Estimation and Hypothesis Testing}
\author{Daniel Fraiman}
\affiliation{Departamento de Matem\'atica y Ciencias, Universidad de San Andr\'es, Buenos Aires, Argentina}
\begin{abstract}
Based only on the information gathered in a snapshot of a
directed network, we present a formal way of checking if the
proposed model is correct for the empirical growing network under
study. In particular, we show how to estimate the attractiveness,
and present an application of the model presented in~\cite{dani}
to the scientific publications network from the ISI dataset.
\end{abstract}
\maketitle

The study of networks has attracted many scientists during the
last decade. The most studied directed growing networks are the
WWW network~\cite{krapi}, where each node represents a web page
and the hyper-links (references to other web pages) represents
the directed edges or links, and the scientific papers
network~\cite{redner}, where each paper is a node, and its
references the directed links. Many of the theoretical approaches
try to determine whether a particular model provides a ``good
description'' of the dynamics of the empirical growing network
under study, paying special attention to the stationary state.
Empirical growing networks shows that this state is characterized
by the fact that it has a degree distribution with a power law
tail, where the degree of a vertex is defined as the total number
of its connections, and a large clustering coefficient, which is
a measure of how much connected are the neighborhoods nodes of
random selected node. These two descriptive statistical measures
characterized in good way the network topology.

\begin{figure}
\begin{center}
\includegraphics[height=0.16\textwidth]{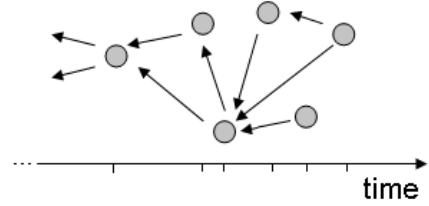}
\end{center}
\caption{Scheme of the growing network model. As time evolves new
nodes with $D_{out}$ number of out-links appear, these ones
attaches to the existing nodes. In this example $D_{out}=\{1,2\}$.
}\label{figrow}
\end{figure}

Typically, the way of checking if a model mimics the real growing
network (once the clustering coefficient is near the empirical)
is comparing the limit degree distribution with the empirical
degree one, paying special attention in the
 tail of the distribution. In~\cite{dani} it was shown that this measure is not a good one, since many
 different models can give the same tail (typically  scale invariant). This way, two other informal
 checks where suggested~\cite{dani} in order to have a first idea whether the model works well.
 The fist one is based on the relation between the variance of the out-degree random variable,
 $D_{out}$ and the in-out degree  covariance, $Cov(D_{in},D_{out})=E(D_{in}D_{out})-E(D_{in})E(D_{out})$.
 Where $D_{in}$ is the in-degree of a randomly selected node.
  The second informal check is based on the
  relation between the tail of the in degree distribution
  and the tail of the out degree one. In~\cite{dani} these two relation
  were studied in detail for the growing model shown in Fig.~\ref{figrow}.

In this letter, we try to go further this informal checks
proposing a way to test whether a particular model describe well
the empirical directed network. Besides, we test the model
presented in~\cite{dani} to see if it describe the scientific
citation network.

Let us now describe the growing network process
(Fig.~\ref{figrow}) presented in~\cite{dani}: as time evolves new
nodes with $D_{out}$ number of out (directed) links appear, which
connect to the existing nodes according to some probability law
(uniform, preferential linking, etc.). $D_{out}$ is a random
variable (or the number of out-links of a randomly selected node)
with an arbitrary distribution, $P(D_{out}=j)=p_j$ with $j\in N$.
For this model, the limit (in/out) degree distribution
($\nu_{in}^k$,$\nu_{in}^k$ and $\nu_{in,out}^{k,j}$) was computed.
 Under preferential linking with attractiveness they get:
\begin{equation}\label{attrac}
\begin{aligned}
& \nu_{out}^{k}=p_k \quad  \quad \quad \quad \quad \quad \quad \quad \quad \quad \quad \quad \quad \quad \quad  (a)\\
& \nu_{in}^{k}=\overset{\infty}{\underset{j=1}{\sum}}p_j\frac{\Psi(j+k+A,3+\delta)}{\Psi(j+A,2+\delta)} \quad \quad \quad  \quad \quad \   (b)\\
& \nu^{k}_{deg}=\Psi(k+A,3+\delta)
\overset{k}{\underset{j=1}{\sum}}\frac{p_j}{\Psi(j+A,2+\delta)} \quad (c)\\
& \nu_{in,out}^{n-k,k}=\nu_{deg,out}^{n,k}=\frac{\Psi(n+A,3+\delta)}{\Psi(k+A,2+\delta)}p_{k}\quad \ \quad (d) \\
\end{aligned}
\end{equation}
where $\Psi(a,b)\equiv
\frac{\Gamma[a]\Gamma[b]}{\Gamma[a+b]}=\int^1_0t^{a-1}(1-t)^{b-1}dt$
, $E_o=\overset{\infty}{\underset{k=1}{\sum}}kp_k$, and
$\delta=A/E_o$.
One advantage of explicitly knowing the dependence of the
$\nu_{in}^{k}$, $\nu_{deg}^{k}$ and $\nu_{in,deg}^{j,k}$
distributions as functions of the out-degree distribution is that
 it is now possible to estimate the attractiveness using just the
 empirical marginal distributions. Let us suppose that we are given
a growing network with a large number of nodes,  whose
distribution we assume close to the limit measure. We take a
snapshot and from the information gathered in the picture we
estimate: 1) the out-degree distribution ($p_k$) by its empirical
law, $\widehat{p}_k=\frac{X_{out}^k}{\underset{k\in N}{\sum}
X_{out}^k }$,  where
 $X^k_{out}$ is the number of nodes with $k$ out-links at the time the picture was taken,
  2) similarly, the in-degree distribution ($\nu_{in}^k$) by
 $\widehat{\nu}_{in}^k=\frac{X_{in}^k}{\underset{k\in N}{\sum} X_{in}^k}$,
 and 3) $E_o$, by $\widehat{E}_o=\underset{k \in
   N}{\sum}k\widehat{p}_k$. On the other hand,
 for each model, $\nu_{in}^k$ be can also estimated from the computed limit in-degree distribution
 (eq.~\ref{attrac} (b) for the model under study).
 In this case, we define $\widetilde{\nu}_{in,A}^k$ as
 $\nu_{in}^{k}$ of eq.~\ref{attrac} (b) after replacing $p_j$ by $\widehat{p}_j$
  and $\delta$ by $A/\widehat{E}_o$. Evidently, $\widetilde{\nu}_{in,A}^k$
   depends on $A$, and the proposed consistent attractiveness
   estimator $\widehat{A}$ becomes:
\begin{equation}\label{asombr}
\widehat{A}=\underset{A \in R_{\geq 0}}{argmin}\{\underset{k \in
N}{max}
|F_{\widetilde{\nu}_{in,A}}(k)-F_{\widehat{\nu}_{in}}(k)|\}
\end{equation}
where $F_{\mu_{in}}(k)$ is the cumulative distribution,
$F_{\mu_{in}}(k)=\overset{k}{\underset{j=0}{\sum}}\mu_{in}^j$.
$\widehat{A}$ is a minimum-distance (for the distribution)
estimator, but there exist many other methods for estimating $A$
such as maximum likelihood or moment method. The advantage of the
first method is discussed below. Alternatively, the $A$ estimation
can be carried out using the degree and the out-degree information
($\widehat{A}=\underset{A \in R_{\geq 0}}{argmin}\{\underset{k \in
N}{max}|F_{\widetilde{\nu}_{deg,A}}(k)-F_{\widehat{\nu}_{deg}}(k)|\}$),
or the joint information. Clearly, an estimator based in the
joint distribution is better since there exist different joint
distributions with the same marginal ones. 
\begin{figure}
\begin{center}
\includegraphics[height=0.4\textwidth]{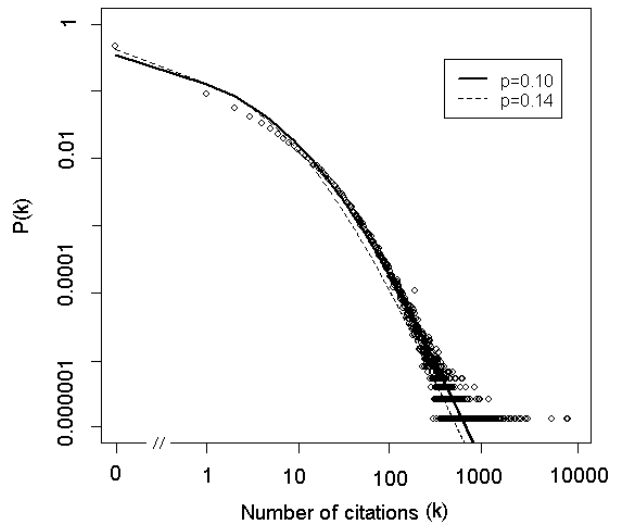}
\end{center}
\caption{Citation distribution for all papers published in 1981
(from the ISI) cited between 1981 and 1997. The theoretical
citation (in-degree) curves are calculated by eq.~\ref{attrac} (b)
assuming that the out-degree distribution is geometric,
$p_k=p(1-p)^k$ for $k\in N_0$. The solid line correspond to
eq.~\ref{attrac} (b) replacing $A$ by 0 and $p$ by $0.1$, the
dashed one correspond to $A=0$ and $p=0.08$.}\label{figpr}
\end{figure}
 Now, we test the model to see if it mimics the
 scientific citation network. The model
 (Fig.~\ref{figrow}) seems to have all the real ingredients of the
 real dynamics of the growing network. Each node represent a scientific publication
 and the directed links the citations. The main point that can be
 criticized is the attachment probability law: in the model it
 depends on the degree (number of papers that cite a random selected paper plus the number of citations of this article) of each node. We discuss another attachment law below.
  Fig.~\ref{figpr} shows the citation distribution of all
 scientific publications in 1981, from the ISI dataset, which were cited between
 1981 and 1997 (see~\cite{redner}). Clearly, this represent the in-degree distribution for a growing network process. Unfortunately the out-degree distribution ($p_k$), the number of cites in a randomly selected paper, has not been reported,
 making a plug-in (see eq.~\ref{attrac}) approach to test the growing model impossible.
 Nevertheless, we adopt the following strategy: we assume a geometric
  out-degree distribution $p_k=p(1-p)^k$ with $k\in N_o$,
  a preferential linking with attractiveness attachment law, and
  estimate $A$ and $p$ by eq.~\ref{asombr}. Clearly,
  the empirical out-degree distribution may not fall in any parametric
  family, however a good estimated in-degree distribution will be a very
useful result, since the in-degree distribution is a theoretical
calculation based on the out one. The T statistic achieves
   its minimum when $A=0$ and $p=0.14$ (see dashed line).
   If we are less ambitious, and disregard the first values of the
   distribution, but require a good match from the $10^{th}$
   citation onwards, we obtain that $p=0.1$ and $A=0$ work remarkably well.
   Note that the theoretical curve (solid line) is extremely similar to the empirical one
   in almost the whole range of the probability. Moreover,
   in this case the mean out-degree ($E_o$) is 9, which seems to be
   a good guess for the average number of cites for all the scientific publications.

\begin{figure}
\begin{center}
\includegraphics[height=0.40\textwidth]{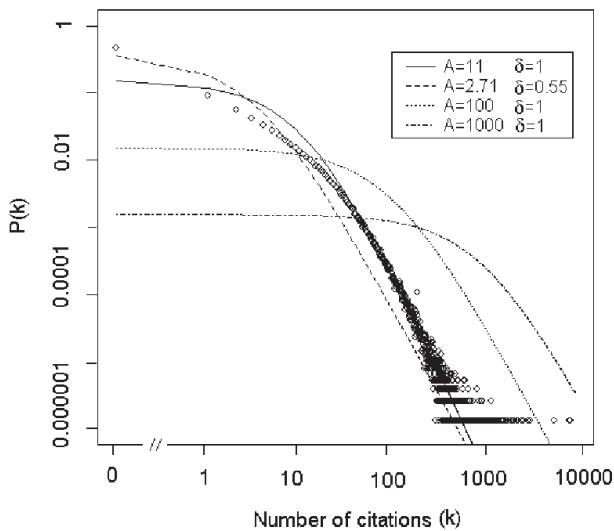}
\end{center}
\caption{Citation distribution for all papers published in 1981
(from the ISI) cited between 1981 and 1997. The theoretical
citation (in-degree) curves are calculated by eq.~\ref{attrac2} (b).
The solid line correspond to eq.~\ref{attrac2} (b) replacing $A$ and $delta$
by different values shown in the legend.}\label{figpr2}
\end{figure}

 As we discussed before, the attachment probability law can be
 criticized.  Perhaps in a better model this law must depend on the in-degree (not on the degree) of each node.
 For this case, using the property 1 introduced in~\cite{dani},
  it is very easy to compute the limit distributions:
\begin{equation}\label{attrac2}
\begin{aligned}
& \nu_{out}^{k}=p_k \quad \quad \quad \quad  \quad \quad \quad \quad \quad \quad \quad \quad \quad \quad \quad \quad  (a)\\
& \nu_{in}^{k}=\frac{\Psi(k+A,2+\delta)}{\Psi(A,1+\delta)} \quad \quad \quad \quad \quad \quad \quad \quad  \quad \quad \   (b)\\
& \nu^{k}_{deg}=\frac{1}{\Psi(A,1+\delta)}
\overset{k}{\underset{j=0}{\sum}}p_j \Psi(k-j+A,2+\delta) \quad (c)\\
& \nu_{in,out}^{j,k}=\nu_{deg,out}^{j+k,k}=\nu_{in}^{j}\nu_{out}^{k}\quad  \quad  \quad \quad \quad \quad \  \quad \quad (d) \\
\end{aligned}
\end{equation}
where $k, j \in N_0$. This case is specially easy to solve because, for a randomly selected node,
 the number of out-links ($D_{out}$) and the number of in-links ($D_{in}$) are independent random variables
 ($\nu_{in,out}^{k,j}=\nu_{in}^{k}\nu_{out}^{j}$). Note that for this attachment law
 the attractiveness must be strictly positive, since
 if $A=0$ we get that the limit in-degree probability is $v^k_{in}=\delta_{k=1}$.
 This result is easy is to understand: new papers appear but they can not cite ($A=0$)
 papers without previous citations, and in this way
the scientific network will be formed by almost all papers with zero citations and only a few very cited.
Clearly, in the limit this goes to a delta function. Moreover, for $k>>1$,
 $\nu_{in}^{k}$ behaves as $k^{-(2+\delta)}$. Now, in the same way as before,
 we will find $A$ and $\delta$ such that $T$ is minimum.
  This happens for $A=2.71$, and $\delta=0.55$ where $T=0.108$
  (see Fig.~\ref{figpr2}). If we are less are less ambitious, and disregard the first values
  of the distribution, but now require a good match from the $30^{th}$
   citation onwards, we obtain that $\delta=1$ and $A=11$ match well the tail.

  Although we know that for both attachment probability
  laws presented the power exponent of the tail of
  the in-degree distribution grows linear with $A$ ($k^{-(3+A/E_o)}$ or $k^{-(2+A/E_o)}$),
   we do not have much intuition of what is the role of attractiveness over the in-degree
distribution. In order to try to understand this, in
Fig.~\ref{figpr2} we show $\nu_{in}^k$ for different values of
the attractiveness. The role of the attractiveness is to make
flatten the first values of the in-degree distribution. The same
type of behavior is observed for the case where the
  law of attachment depends on the degree of the node (data not shown).
  Once we have the figure, the interpretation is immediate since for greater
  $A$ the difference between selecting a node with
  let say 2 citations will be very similar to the one with 1 or 3 citations
  ($\pi_{in}^k=(k+A)\nu_{in}^k/(E_{o}+A)$).

In general a growing network model can be separated in two parts:
A) the model in its own (e.g. in each temporal step a new node
with $D_{out}$ number of links is aggregated) , and B) the
attachment probability law for the new links. We have previously
shown two models that differ only on part B). It is important to
check whether a proposed model (A+B) describes well the empirical
data. If we are convinced that the part A) is correct for
describing an empirical growing network, we can test if the
attachment probability law is for example preferential linking
(or uniform). In the case where you are convinced that the
attachment law has some known law (part B) correct), we can test
if the part A) of the model is adequate. In the other case, it is
also possible to test both hypothesis (the model), but if $H_o$
is rejected we can not determine which (A or B or both) is the
incorrect.

One advantage of the proposed estimator in eq.~\ref{asombr} is
that it is now very easy to test any of the hypothesis previously
mentioned. For example $H_o$: the real growing network has an
underlying link attachment law that is preferential with
attractiveness. This hypothesis can be tested~\cite{kolmo} with
the usual statistic, $T=\underset{k \in
N_o}{max}|F_{\widetilde{\nu}_{in,\widehat{A}}}(k)-F_{\widehat{\nu}_{in}}(k)|$.
This test is one of the main result presented here.

Another important issue is to be able to rank the models. For
example, in~\cite{tipos_attach} a nice growing model was proposed
for the WWW dynamics. In this model, with probability $p$ a new
node with only one out-link is aggregated, or a new directed link
from an existing node is created (with probability $1-p$). This
is the part A) of the model, and the following constitute the
part B) of the model. The new link from the new node is attached
by preferential linking with attractiveness for the in-degree,
$\pi_{in}^{k}= \frac{k+A}{E_{in}+A}\nu_{in}^k$
(in~\cite{tipos_attach} they use $\lambda$ instead of $A$ and
works with rates and not with probability). And the selection of
the new created link, has two independent events: 1) the
selection, by preferential linking with attractiveness for the
out-degree, of the origin, 2) and the selection of the target by
preferential linking with attractiveness (with a different
parameter from the previous one) for the in-degree. An
alternative model can be the one proposed for the scientific
network (see Fig\ref{figrow}), where now the nodes represent the
web pages and the links the hyper-links.  Clearly, both models
have some weak points. In the first one~\cite{tipos_attach},
people can not put more than one hyper-link when they are
constructing their own web page (later some new links can
appear). The second model (Fig~\ref{figrow}) do not have this
inconvenient, but is ``static'', once the hyper-links are fixed
they can not be changed.
 Probably, a mixed model between both be more realistic.
 How to compare or rank these model is a relevant question in
 order to approach to the ``real model''. There exit many
 statistical measures that do this job, in particular there is an
 extent bibliography for Hidden Markov chain problems, and also for linear regression models. We propose
 a very simple ranking variable that is the value of $T$ using the
 (cumulative) joint distribution. The best model is the one that has the
 minimum value of T. Much work must be done to understand how must
 be the penalization (if it is necessary) for models with many parameters.

In summary, we discussed: 1) how to estimate model parameters,
showing an application to the scientific publications network,
and 2) a way of checking whether a proposed model is correct,
based on the limit (joint) in and out degree distribution.
This way, the results presented here shed some light on the
problem of estimating the underling attachment law, ranking
models, and test models in a general way.


\end{document}